\documentclass{mn2e}
\usepackage{psfig}

\def\ltsima{$\; \buildrel < \over \sim \;$}
\def\lsim{\lower.5ex\hbox{\ltsima}}
\def\gtsima{$\; \buildrel > \over \sim \;$}
\def\gsim{\lower.5ex\hbox{\gtsima}}
\def\mes{M\'esz\'aros}

\begin{document}

\title[Photon scattering in GRBs]
{The role of photon scattering in shaping the lightcurves and spectra
of $\gamma$-ray bursts}

\author[Lazzati]
{Davide Lazzati\\
Institute of Astronomy, University of Cambridge, Madingley Road,
Cambridge CB3 0HA, England \\
{\tt e-mail: lazzati@ast.cam.ac.uk}
}

\maketitle

\begin{abstract}
We analyze the power spectra of the lightcurves of long gamma-ray
bursts, dividing the sample in bins of luminosity, using the recently
discovered variability-luminosity correlation. We find that the value
of the variability parameter strongly correlates with the frequency
that contains most of the power in the burst comoving frame. We
compute the average power spectra in luminosity bins. The average
power spectrum is well described by a broken power-low and the break
frequency is a function of the variability parameter, while the two
slopes are roughly constant. This allow us to conclude that scattering
processes do not play a relevant role in modelling the lightcurves. We
finally discuss in which conditions scattering may still play a
relevant role in shaping the spectra of GRBs.
\end{abstract}


\section{Introduction}

The study of the BATSE $\gamma$-ray bursts (GRBs) lightcurves has
recently gained new interest thanks to the discovery of the
variability-luminosity (Fenimore \& Ramirez-Ruiz 2000; Reichart et
al. 2001) and lag-luminosity (Norris et al. 2000) correlations (see
also Chang, Yoon \& Choi 2002). These enable to assign a tentative
redshift to BATSE GRBs, allowing to perform spectral and temporal
analysis of the lightcurves in the burst comoving frame, where their
properties are more closely linked to the physics of the burst
itself. As an example, the use of these correlations enabled the
possible discoveries of an evolution of the luminosity function with
redshift (Lloyd-Ronning, Fryer \& Ramirez-Ruiz, 2002) and of a
correlation between the peak photon frequency with the luminosity of
the GRB (Lloyd-Ronning \& Ramirez-Ruiz 2002).

The variability-luminosity correlation predicts that the peak
luminosity of a burst is correlated to its degree of variability,
whose operational definition is related to the normalized variance, or
the root mean square of the deviations from a smoothed version of the
light curve. More variable lightcurves have larger luminosities. The
lag-luminosity correlation is instead based on the measure of a
temporal lag between the detection of high energy and low energy
photons. It is found that the more the burst is luminous, the smaller
is the lag. Besides being extremely useful tools, these correlations
also call for an explanation of their origin. Several possible
interpretations have been discussed in the literature. In particular,
it is shown that an underlying correlation between the isotropic
equivalent luminosity and the Lorentz factor of the fireball can
explain both the correlations (Ramirez-Ruiz \& Lloyd-Ronning 2002;
Kobayashi, Ryde \& MacFadyen 2002; Salmonson 2000; \mes~et
al. 2002). In some of these works, however, a significant role is
attributed to the modification of the temporal properties of the light
curves as a consequence of scattering by cold or hot electrons (see
also Panaitescu, Spada \& \mes~1999; Spada, Panaitescu \&
\mes~2000). The role of scattering seem to be strengthened by the need
of explaining the peak photon frequency vs. isotropic equivalent
luminosity correlation (Lloyd-Ronning \& Ramirez-Ruiz 2002), which is
not naturally predicted in the internal shock scenario (Ghisellini,
Celotti \& Lazzati 1999).

In this paper, we concentrate on the variability luminosity
correlation, using power spectra as a diagnostic to investigate
several related issues. First (\S 2), we compute the dominant
frequency of the power spectrum for each GRB in our sample of 220
lightcurves. We find that this dominant frequency strongly correlates
with the variability measure. We then divide the sample in bins of
variability and compute the average power spectrum. We find (\S 3)
that the power spectra are self-similar (Beloborodov, Stern \&
Svensson 1998; 2000; hereafter B98 and B00; Chang \& Yi 2000) in all
the variability bins, with a break at a frequency that correlates with
the variability parameter. No sign of a cut-off in the spectrum at
large frequencies is observed. We (\S 4) develop a shot-noise model
for the lightcurve, taking into account the effect of scattering in
Fourier space (\S~5). In \S 6 we discuss our results, showing how
scattering processes cannot be responsible for the
variability-luminosity correlation, and also constraining the regions
of the parameter space where photon scattering on cold electrons may
imprint detectable signatures on the photon spectra of GRBs without
leaving a detectable trace in their power spectra.

\section{Data Analysis}

\begin{figure}
\psfig{file=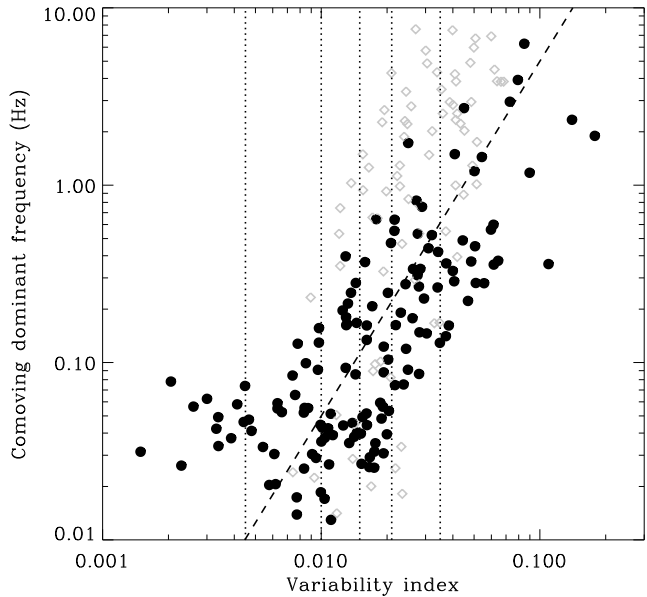,width=0.48\textwidth}
\caption{{The dominant frequency in the burst comoving frame vs. the 
burst variability parameter. Solid points show the bes class events
(see Sect. 2 for details) while gray symbols show the lower class
events. A correlation is clearly evident. The dashed line shows a
relation $\nu_d\propto{V}^2$ as a guideline for the first class
events. Vertical dotted lines show the boundaries of the subclasses in
which the sample has been divided.}
\label{fig:cor}}
\end{figure}

We have adopted the whole sample of GRBs for which Fenimore \&
Ramirez-Ruiz (2001) computed the variability parameter and a tentative
redshift. For each lightcurve, we have computed the power spectrum
(hereafter PSD) as $PSD(\nu)=|{\tilde{f}}(\nu)|^2$, where
${\tilde{f}}(\nu)$ is the Fourier transform of the
lightcurve. Lightcurves were binned to 64 ms resolution and considered
from the BATSE trigger time for a total duration of $t=3*T_{90}$,
where $T_{90}$ is the time containing $90\%$ of the total burst
emission.  A cubic function was fitted to the time interval before and
after the one considered above in order to remove the background. The
effect of background removal was however weak, affecting only the
smallest frequencies.

The dominant frequency $\nu_d$ was detected as the frequency at which
the power per unit decade frequency is maximized (something similar to
the peak of the $\nu\,F(\nu)$ photon spectrum), i.e. the maximum of
the function $\nu\,PSD(\nu)$. We adopted this definition since it does
not involve any fitting procedure and is therefore the most
objective. The definition of the dominant frequency must confront,
however, with two problems. First, we must consider that the time
series is limited (see below), secondly the presence of a white noise
component, which is due to the Poisson noise of the lightcurve (which
is inevitably affected by photon count statistics). This component
results in a flat power spectrum (see e.g. Leahy et al. 1983 and
references therein) which produces a false peak at the highest
considered frequency when the $\nu\,PSD(\nu)$ function is considered
for lightcurves in which the signal is weaker than a certain
threshold. In order to avoid this contamination of spurious peaks, the
spectra were all visually inspected, and every time the peak frequency
was in the flat part of the PSD, the considered frequency range was
shrunk in order to recompute the maximum in a shorter frequency range,
where the signal-to-noise ratio is larger. This procedure,
unavoidably, reduces the objectivity of the analysis, and for this
reasons the lightcurves in which the peak selection was modified by
hand were flagged. In Fig.~\ref{fig:cor} we show the comoving frame
dominant frequency vs. the variability parameter for all the best
class lightcurves, in which human intervention was either not required
or clearly unbiased. A clear correlation is present. The correlation,
according to the correlation coefficient statistics, is highly
significant, with a vanishing probability of being spurious
$P=3\times10^{-25}$. It may be argued that the correlation is
artificially made by a correlation between the redshift and the
variability (which is indeed present in the data). In order to test
for this to be true, we computed the above probability also for the
observed frame frequency. Again, a strong correlation is found, with
probability of being spurious $P=8\times10^{-12}$. The lower class
bursts, in which human intervention was necessary and possibly biased,
or in which the dominant frequency was the lowest one, are plotted in
light gray in Fig.~\ref{fig:cor}. It is possible to see how these
points do not invalidate the correlation, even though the dominant
frequencies tend to be, as expected, larger than for the best
cases. In the following we will perform our analysis only on the
subsample of 143 best lightcurves shown in Fig.~\ref{fig:cor}, even
though extending it to the whole sample does not affect significantly
the conclusions. It is worth stressing that such a correlation is not
entirely surprising, since it simply says that the variability
parameter has something to do with the mean frequency over the
PSD. What is relevant, as we are going to discuss in much more detail
in the following, is that this correlation tells us that a change in
the variability parameter $V$ reflects a change in the whole PSD
rather than the suppression of high (or low) frequency power.

\begin{figure*}
\psfig{file=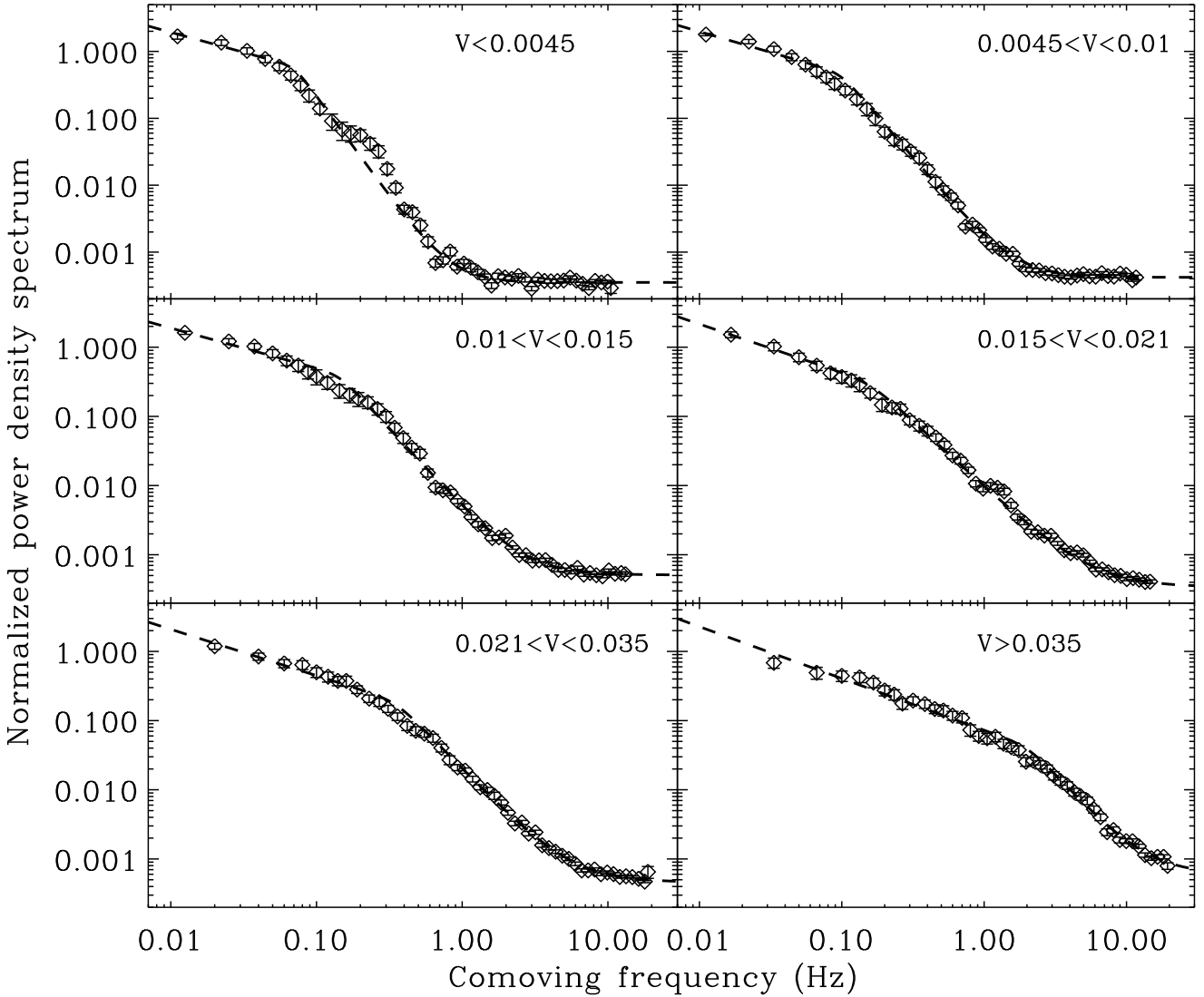,width=\textwidth}
\caption{{Average power spectra of burst in the six variability subclasses 
defined in the text (see also Fig.~\ref{fig:cor}). The spectra have
been binned in frequency in order to have a constant number of points
per logarithmic unit frequency. The gray line shows the best broken
power law (plus constant) fit. Error bars are derived from the
dispersion of the sample.}
\label{fig:ave}}
\end{figure*}

When computing the PSD of an experimental time series, one has to take
into account all the effects due to the data binning and to the finite
duration of the data. In particular for GRBs, this second issue is
particularly tricky, since the data are truncated not by hand, but by
the intrinsic duration of the event. Consider an infinite function
$f_\infty(t)$. The binning process can be modelled as the convolution
with a square filter $b(t)$ and the multiplication with a lattice
$s(t)$ (a series of Dirac $\delta$ functions) with spacing equal to
the full width of the square filter. The truncation of the data is
represented as the multiplication with a window function $w(t)$ which
is usually assumed to be the characteristic function of the considered
interval, but may have different shapes in the case of GRBs (see
below). Thanks to the properties of multiplication and convolution in
Fourier space, the Fourier transform of $f$ can be written as (Lazzati
\& Stella 1997; hereafter LS97):
\begin{equation}
{\tilde{f}}(\nu)=\left\{\left[
{\tilde{f}}_\infty(\nu)\ast{\tilde{w}}(\nu)
\right]\,{\tilde{b}}(\nu)\right\}\ast{\tilde{s}}(\nu)
\label{eq:fft}
\end{equation}
where the symbol $\ast$ indicates a convolution. Let us analyze the
implications of Eq.~\ref{eq:fft} from right to left. The Fourier
transform of the lattice of spacing $\delta t$ (in our case 64 ms) is
the reciprocal lattice, i.e. a lattice with spacing $2\pi/\delta
t\sim100$~s. This term is important if periodicities are present in
the lightcurve, since it gives rise to the well known phenomenon of
aliasing. In our case, since the power spectra are monotonically
decreasing with frequency, it is irrelevant. The Fourier transform of
the binning function has a form (see, e.g., LS97)
\begin{equation}
{\tilde{b}}(\nu) \propto {{\sin\left(\pi\,\delta t\,\nu\right)}
\over{\pi\,\delta t\,\nu}}
\label{eq:sinc}
\end{equation}
This is the most dangerous term, since it introduces a break at a
frequency $\nu\sim1/\delta t\sim16$~Hz in the observer frame. The
largest of our dominant frequencies, in the observer frame, is
$\max(\nu_d)=0.5$~Hz. We conclude that this effect does not play a
relevant role in our analysis. Finally, the effect of the window
function is to smooth the observed PSD. In the most common case of a
square function, the smoothing kernel is similar to the function in
Eq.~\ref{eq:sinc} but with a smaller width. Again, this term is
important is small scale features overlaid on the spectrum are
concerned. For any reasonable shape of the window function we can
ignore its effect.

We therefore conclude that the correlation of Fig~\ref{fig:cor} is
real. It is worth noting that the typical frequencies, that seem to
dominate the determination of the variability parameter $V$, are
small, even in the comoving frame. This suggests that it is the
$\sim1$ second variability which is physically linked to the
luminosity of the GRB rather than its shorter time scale fluctuations
(see also below). There is also a suggestion, in the data, that the
very small $V$ bursts belong to a different family in terms of their
temporal properties. We will discuss this in the following, finding
that the average power spectra seem to be more consistent with a
continuation of the average spectral properties rather than with a
different subclass.

\begin{table}
\begin{center}
\begin{tabular}{c||cccc}
$\langle V\rangle$ & $\alpha_1$   & $\alpha_2$  & $\nu_0$ (Hz) & Prob($\sigma$) 
\\ \hline \hline
0.0032           & $0.6\pm0.18$ & $3.\pm0.15$ & $0.071\pm0.016$ & 7.2\\
0.0077           & $0.62\pm0.1$  & $2.5\pm0.1$  & $0.098\pm0.017$  & $>8.3$\\
0.0125           & $0.58\pm0.09$ & $2.3\pm0.08$  & $0.145\pm0.03$ &   $>8.3$\\
0.018            & $0.7\pm0.13$  & $1.9\pm0.06$  & $0.16\pm0.04$   & 8.1\\
0.027            & $0.67\pm0.09$ & $2.16\pm0.07$ & $0.35\pm0.07$  & $>8.3$\\
0.053            & $0.75\pm0.08$ & $2.22\pm0.3$  & $2.0\pm0.6$    & 8.2
\end{tabular}
\end{center}
\caption{{Results of the fit of the average power spectra in 
Fig.~\ref{fig:ave}. The rightmost column show the significance for the
existence of the break.}
\label{tab:uno}}
\end{table}

\section{Average power spectra}

We have divided the sample of 143 GRBs in six bins of the variability
parameter $V$. Each bin contains $\sim 25$ lightcurves, with the
exception of the smallest $V$ bin, in which only 12 lightcurves are
contained. This difference is due to the need of keeping these bursts
isolated since they seem to belong to a different class. The
lightcurves, background subtracted, were normalized to contain the
same numbers of photons and the power spectra averaged at the same
comoving frequency. The errors were derived from the dispersion of the
sample.

This process is similar to what was done by B98 and B00. Their average
power spectrum presents however three fundamental differences. First,
they performed the average at the same observed frequency, instead of
the comoving frequency. This was due to the fact that redshift
estimates were not available at that time. Secondly, they decided to
normalize their lightcurves to the same peak photon luminosity rather
than to the same photon fluence. We decided to use this second
normalization since the scatter in the sample of spectra was lower in
this case (see B98 for a discussion). Finally they built the average
of all the lightcurves, while we subdivide our sample in bins of
variability.

The resulting average power spectra are shown in
Fig.~\ref{fig:ave}. The variability parameter increases from left to
right and from top to bottom. The spectra seem to be made by a broken
power law, sinking into the white noise at large frequencies. It is
worth to note that our normalization criterion does not produce
uniform white noise values. For this reason the subtraction of this
component from the average spectra is not possible, and it is also
dangerous to investigate the frequency region where the average
spectra are dominated by white noise.

We fit the average spectra with smoothly a broken power-law function
of the form
\begin{equation}
PSD(\nu)= {{2F_0}\over{
\left[\left({{\nu}\over{\nu_0}}\right)^{2\alpha_1}+
\left({{\nu}\over{\nu_0}}\right)^{2\alpha_2}\right]^{1/2}
}}+K
\end{equation}
where $\alpha_1$ is the slope at frequencies smaller than the break
$\nu_0$, $\alpha_2$ is the slope at larger frequencies and $K$ is a
constant that takes into account the contribution of the white
noise. $F_0$ is the value of the PSD at the break frequency.

\begin{figure}
\psfig{file=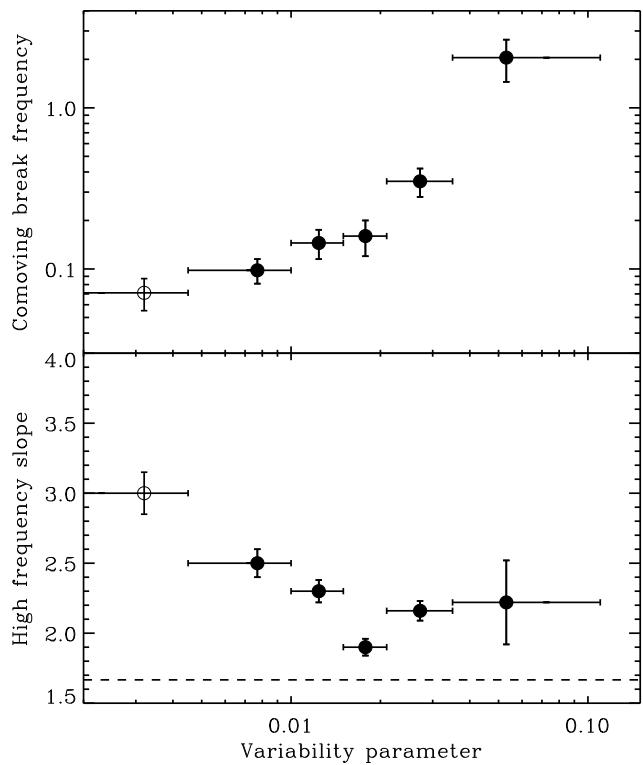,width=0.48\textwidth}
\caption{{The break frequency and high energy slopes of the average power 
spectra versus variability. The upper panel shows the knee frequency
as a function of variability. The first point (open circle) is for the
lowest variability class. The lower panel shows the high frequency
spectral index versus the variability parameter.}
\label{fig:res}}
\end{figure} 

The results of the fit are reported in Tab.~\ref{tab:uno} and in
Fig.~\ref{fig:res}. Several interesting remarks can be derived from
the results. First, all the power spectra are well fit by the
model. Moreover, the slopes of the broken power-law before and after
the break are roughly consistent with being constant, with the break
frequency $\nu_0$ being the only quantity that evolves with the
variability parameter $V$. A marginal evidence of evolution of the
high frequency power law index is also present, especially if the
smallest variability bin is included in the sample. A second important
remark is that the high frequency slope is larger than what found by
B98.  This difference is not due to the redshift correction nor to
the different normalization. In fact, if we average among them the six
power spectra of Fig.~\ref{fig:pds}, we recover with good accuracy the
$-5/3$ slope in B98. It looks therefore like that the slope can be
attributed to the convolution of the break frequencies in different
variability bins. Finally, even though the lowest variability bin
values are different from the rest of the sample, it is not possible
to define it as a separate class given the present data. The apparent
segregation of the dominant frequencies in Fig.~\ref{fig:cor} may be
due to the fact that the dominant frequencies of the very low
variability bursts approach the lower cut in the analyzed frequencies,
and include therefore an additional noise term.

\section{Shot noise model}

In the internal shock model for GRBs (Rees \& \mes~1994) the
lightcurve is modelled as the random superposition of a number of
pulses with similar shape stretched in shape and scaled in luminosity
according to some prescription for the ejection of shells by the inner
engine (Kobayashi Sari \& Piran 1997; Panaitescu et al. 1999; Spada et
al. 2000, 2001; Ramirez-Ruiz \& Lloyd Ronning 2002). This lightcurve
is similar to the shot noise model supposed to play a relevant role in
the red-noise component observed in the power spectra of X-ray pulsars
(LS97 and references therein).

Consider a normalized pulse of shape $p(t-t_0,\tau_r)$, with
rise time $\tau_r$ and unit decay time and fluence peaking a time
$t_0$. A GRB lightcurve can be described as a random superposition of
$N$ such pulses:
\begin{equation}
lc(t)=\sum_j f_j\,p\left({{t-t_j}\over{\alpha_j}},\tau_r\right)
\label{eq:lci}
\end{equation}
where the pulse fluence $f_j$, the stretching factor $\alpha_j$ and the
peak time $t_j$ are randomly selected according to some prescription
(for example simulating the hierarchical shock evolution of an
inhomogeneous flow). The Fourier transform of the above
equation can be written (up to a normalization factor) as:
\begin{equation}
{\tilde{lc}}(\omega) \propto \sum_j
{\tilde{f}}(\alpha_j\omega)e^{-i\omega t_j}
\label{eq:ft}
\end{equation}
When we compute the average power spectrum of a sample of lightcurves,
under the assumption that each lightcurve is a random realization of
the same underlying process, we compute the ensemble average of the
square of the modulus of Eq.~\ref{eq:ft}. This, if the peak times of
the pulses are not correlated with their properties and are uniform in
time\footnote{It is a well known result that the average width of the
pulses does not evolve during the bursts (Ramirez-Ruiz \& Fenimore
2000), but their frequency and/or fluence may be larger in the early
phase of bursts. This correlations takes the form o a non-square
window function in Fourier space, and are not relevant here, as
discussed in~\S~2. Finally, it may be wrong to assume that the pulse
fluence does not depend on the pulse duration (Ramirez-Ruiz \&
Fenimore 2000). Should this assumption be wrong, it would reflect in a
different numeric value for Eq.~\ref{eq:sixt}.}, is given by (LS97):
\begin{equation}
\langle PSD\rangle\propto\langle{\tilde{f}}^2(\alpha\omega)\rangle
\label{eq:apds}
\end{equation}
i.e. the average power spectrum is independent on the time history of
the pulse ejection and on the pulse fluence distribution, but depends
on the distribution of the pulse durations. To understand the effect
of the above conclusion, we consider a simplified exponential pulse
profile:
\begin{equation}
p(t)=e^{-t}\chi_{(0,\infty)}
\label{eq:sp}
\end{equation}
where $\chi_{(a,b)}$ is the characteristic function of the interval
$(a,b)$. In addition, we consider a power-law distribution\footnote{It
has been found that the duration distribution of ``well separated''
pulses is well described by a Log-normal function (McBreen et
al. 1994; Li \& Fenimore 1999). We prefer to adopt a power-law
distribution here since it fits better our average power-spectra and
since our PSDs do include {\it non well separated} pulses.} of
$\alpha$ values between a minimum value $\alpha_m$ and a maximum value
$\alpha_M$:
\begin{equation}
n(\alpha)\propto\alpha^a\chi_{(\alpha_m,\alpha_M)}
\label{eq:alp}
\end{equation}

The power spectrum of the single pulse of Eq.~\ref{eq:sp} is a
Lorentzian function, i.e. (roughly speaking) a constant for
$\nu<1/2\pi$ and a power-law $\nu^{-2}$ for $\nu>1/2\pi$. For the
stretched pulse, the PSD is flat for $\nu<1/(2\pi\alpha)$ and a power
law afterword.  Consider now the average PSD of Eq.~\ref{eq:apds} for
the $\alpha$ distribution defined in Eq.~\ref{eq:alp}. If $a\le-1$ the
PSD will be dominated by the shortest pulse, and will retain a
Lorentzian shape. For $a\ge1$, it will be dominated by the longest
pulse, again retaining the Lorentzian shape. For $-1<a<1$, a new
power-law branch will appear, in the range
$1/(2\pi\alpha_M)<\nu<1/(2\pi\alpha_m)$, with a slope
$\nu^{-(a+1)}$. Since for $a\ge1$ the PSD is dominated by the longest
pulse, here and in the following we will define the ``shortest
relevant pulse'' as the shortest pulse that have an influence in the
PSD shape. This pulse may not be the shortest observed in the
lightcurve. For example, consider a broken power-law distribution of
pulse durations, with $-1<a<1$ at short durations and $a\ge1$ at short
durations. the break in the power spectrum will be related to the
pulse at the break of this distribution. We call this the shortest
relevant pulse.

\begin{figure}
\psfig{file=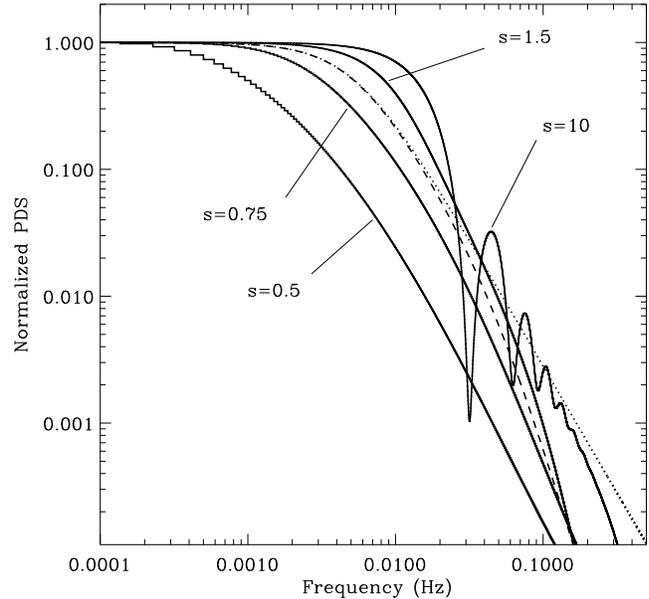,width=0.48\textwidth}
\caption{{Power spectra for stretched exponential pulses as defined 
in Eq.~\ref{eq:str}, for the set of parameters $\tau_r=3$~s,
$\tau_d=30$~s, and the values of $s$ indicated. The dotted line shows
the power spectrum for a single exponential profile while the dashed
line shows the PSD for a double sided exponential profile.}
\label{fig:str}}
\end{figure}

The shape of the fundamental pulse in Eq.~\ref{eq:sp} can be made
more complicated. Consider as an example, a double exponential pulse,
of functional shape:
\begin{equation}
p(t,\tau_r,\tau_d) = {1\over{\tau_r+\tau_s}} 
\left[e^{{t}\over{\tau_r}}\chi_{(-\infty,0)}+
e^{-{{t}\over{\tau_d}}}\chi_{(0,+\infty)}
\right]
\end{equation}
Its PSD can be written as:
\begin{equation}
\left| {\tilde p}(\omega) \right|^2 \propto {{1}\over{
1+\omega^2\tau_d^2\left(\omega^2\tau_r^2+1+\tau_r^2/\tau_d^2\right)}}
\label{eq:pds}
\end{equation}
where $\omega\equiv2\pi\nu$ is the pulsation. If $\tau_r\ll\tau_d$,
Eq.~\ref{eq:pds} can be approximated as a double broken power-law:
\begin{equation}
\left| {\tilde p}(\omega) \right|^2 \propto
\left\{\begin{array}{ll}
\nu^0 & \nu\ll 1/2\pi\tau_d \\
\nu^{-2} & 1/2\pi\tau_d\ll\nu\ll 1/2\pi\tau_r \\
\nu^{-4} & \nu\gg1/2\pi\tau_r
\end{array}\right.
\end{equation}
Again, the ensemble average of Eq.~\ref{eq:apds} can add different
power-law branches if $-2<a<0$. Finally let us consider a stretched
double exponential (Norris et al. 1996) of equation:
\begin{equation}
p(t,\tau_r,\tau_d) \propto
\left[e^{\left({{t}\over{\tau_r}}\right)^s}\chi_{(-\infty,0)}+
e^{-{\left({{t}\over{\tau_d}}\right)^s}}\chi_{(0,+\infty)}
\right]
\label{eq:str}
\end{equation}
For $s<1$ this gives a pulse with a very spiky core and broad wings,
while for $s>1$ the resulting pulse has a square shape. The PSD cannot
be analytically computed. In Fig.~\ref{fig:str} we show the shape of
the PSD for a set of values of $s$, compared to the PSD of the single
and double exponential pulses. For $s<1$ the effect is of smoothing
out the breaks and extending the power-law branch $\nu^{-2}$, while
the effect of $s>1$ is more complex. For $1<s<2$ the break frequency
is moved to larger values, while the slope of the power-law decay is
increased. For $s>2$ the PSD starts to deviate from the simple form,
with the appearance of ``absorptions''. In an ensemble average these
small scale features will be erased and we are left with the envelope,
which yields again a very steep power-law slope.

\begin{figure}
\psfig{file=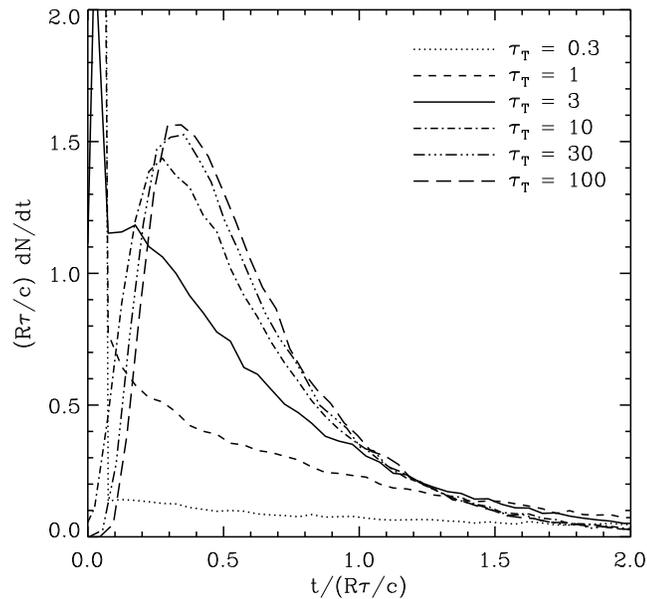,width=0.48\textwidth}
\caption{{Transfer function shapes for a uniform cloud of perfect
scattering centers with radius $R$ and opacity $\tau_T$. Different
line styles show different opacities. For $\tau_T\gg10$ the shape of
the transfer function becomes scale invariant. Each curve is obtained
by a Monte Carlo simulation with $10^5$ input photons.}
\label{fig:ker}}
\end{figure}

\begin{figure}
\psfig{file=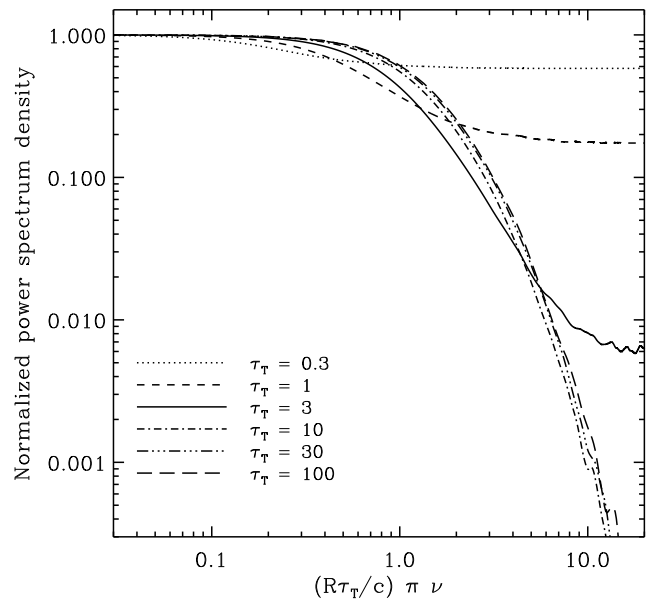,width=0.48\textwidth}
\caption{{Power spectral density (normalized to the value at zero frequency) 
of the transfer functions shown in Fig.~\ref{fig:ker}, as a function
of the dimensionless frequency $(R\tau_T/c)\,\pi\,\nu$. For $\tau_T>1$
the PSDs show a prominent break at the unit value of the dimensionless
frequency.}
\label{fig:pds}}
\end{figure} 

\section{Photon scattering}

It has been proposed that photon scattering may play a relevant role
in shaping both the temporal (Panaitescu et al. 1999; Spada et
al. 2000; Kobayashi et al. 2002; Ramirez-Ruiz \& Lloyd-Ronning 2002)
and spectral (\mes~\& Rees 2000, \mes~et al. 2002) properties of GRB
lightcurves.  The effect of photon scattering has a distinctive
signature in Fourier space, and the power spectrum of the lightcurves
is then the best way in which the importance of scattering can be
evaluated.

Consider a source of photons producing a flash (ideally a Dirac
$\delta$ function in time) in the centre of a cloud of ideal
scattering particles\footnote{Ideal means in this context, that the
photons are never absorbed and that the scattering properties of the
cloud do not depend on the photon energy.}. Even though the central
source emitted a photon impulse, an observer located outside the cloud
would detect a light pulse with a finite duration and a particular
shape. This is due to the fact that different photons make different
paths, finally reaching the observer after being scattered many times
in random directions. The shape of this pulse will be function of the
opacity of the cloud to photon scattering only, up to a scale factor
that takes into account the size of the cloud.

Consider now the central source itself producing a signal with finite
duration and with a given shape $s(t)$. This can be approximated as an
infinite series of spikes, each of them being detected by the outside
observer as a pulse of shape $k(t)$. In mathematical terms, the
observer will detect a signal $S(t)$ given by the convolution of the
original function $s(t)$ times the transfer function $k(t)$:
\begin{equation}
S(t)=s(t)\ast k(t) \equiv \int_{-\infty}^\infty s(t-t^\prime)\,
k(t^\prime)\,dt^\prime
\end{equation}
Thanks to the properties of Fourier transforms, the PSD of the detected
signal $S$ is the product of the spectra of $s$ and $k$.

In the case of photon scattering by free electrons, we consider a
cloud of radius $R$ and uniform density $n$, with opacity
$\tau=R\,n\,\sigma_T$. We neglect the angular dependence of the
Thompson cross section. In Fig.~\ref{fig:ker}, we show the transfer
functions obtained by Monte-Carlo calculations of scattering for a set
of optical depths. For $\tau\le1$ the transfer functions are Dirac
$\delta$ function with a small tail at large times, but for $\tau>1$
the vast majority of photons undergo at least one scattering and the
transfer function becomes much more smooth. For $\tau>10$ the opacity
itself is not a parameter any more but a scale factor:
\begin{equation}
k_{\tau>10,{R\over c}} (t) = {c\over{R\tau}}k_{10,1}
\left({{ct}\over{R\tau}}\right)
\end{equation}

\begin{figure*}
\psfig{file=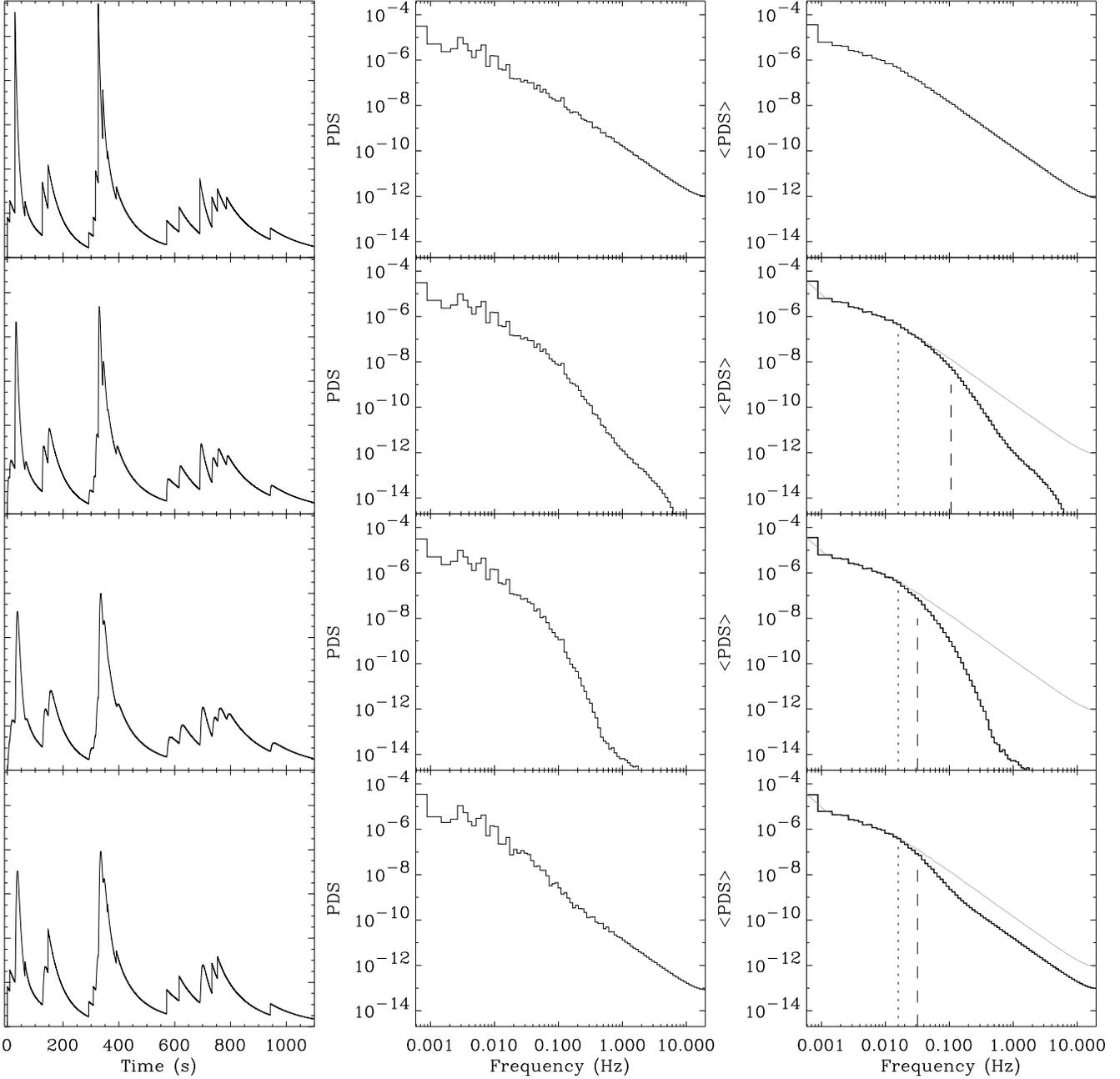,width=\textwidth}
\caption{{Simulated lightcurves and power spectra of a shot lightcurve 
with various degrees of scattering (see text for more details). The
leftmost column shows the a sample simulated lightcurve, whose PSD is
shown in the central column. The rightmost column shows the average
PSD of 100 realizations of such a lightcurve. The first row show the
case with no opacity and scattering. In the second row all the pulses
are smeared by a scattering slab with $R\tau/c=3$. In the third row
all the pulses are smeared by a scattering slab with $R\tau/c=10$. In
the last row, the properties of the slab are held the same, but only
the shortest pulses are smeared while the longest are unperturbed. In
the rightmost column, the vertical dashed line shows the position of
the scattering break, the vertical dotted line shows the position of
the pulse break and the gray line show the unscattered average
spectrum of the uppermost row. All the simulations are made by
single exponential shots {\bf (Eq.~\ref{eq:sp})} and contain 20
pulses. The pulse fluence has a log-normal distribution, while the
pulse duration $\tau_d$ is distributed according to Eq.~\ref{eq:alp}
with $a=-1/2$, $\alpha_m=10$~s and $\alpha_M=250$~s. }
\label{fig:sim}}
\end{figure*}

This behavior is reflected by the shape of the transfer functions in
Fourier space. In Fig.~\ref{fig:pds} we show the PSD of the same
transfer functions shown in Fig.~\ref{fig:ker}. For small opacity, as
expected, the PSD of the transfer function is almost a unit constant,
so that ${\tilde S} (\omega) \simeq {\tilde s}(\omega)$. As the
opacity increase, a pronounced break appears at a frequency
\begin{equation}
\nu_b={c\over{\pi\,R\,\tau}}
\end{equation}
For intermediate opacities, the PSD can be roughly described as a
power-law for $\nu>\nu_b$ while for large opacity the break takes the
form of an exponential cutoff. In conclusion, the effect of photon
scattering is to create a clear break in the power spectrum,
suppressing all the frequencies larger than $\nu_b$.  In
Fig.~\ref{fig:sim} we show the result of a set of simulations that
include scattering at various degrees. The left column shows simulated
lightcurves with a single exponential shot noise model
(Eq.~\ref{eq:sp}) with 20 pulses. The pulse fluence has a log-normal
distribution, while the pulse duration $\tau_d$ is distributed
according to Eq.~\ref{eq:alp} with $a=-1/2$, $\alpha_m=10$~s and
$\alpha_M=250$~s. The top lightcurve has no scattering, while the
second has $\tau=3$ and the third has $\tau=10$. In all cases
$R/c=1$. The last curve has been made by smoothing with $\tau=10$ only
the 7 shortest pulse. This may simulate more closely the effect of
scattering in the internal shock scenario, where shorter pulses are
produced closer to the inner engine, where the opacity of the flow is
larger (see, e.g., Spada et al. 2001 and references therein). The
second column shows the PSD of the lightcurves shown in the first
column, while the third column shows the average PSD of 100 curves
generated according to the same pulse process. The difference between
the scattered and unscattered PSDs is shown by overlaying the
unscattered PSD with a light grey curve.

As mentioned above, the effect of photon scattering in defining
the duration of GRB pulses has been considered in a number of works
(Panaitescu et al. 1999; Spada et al. 2000; Kobayashi et al. 2002;
Ramirez-Ruiz \& Lloyd-Ronning 2002). It is worth mentioning, however,
that the way in which this was done is not completely accurate,
especially when the power spectrum is concerned. In fact, in numerical
works, the duration of a pulse is computed as the sum in quadrature of
the intrinsic duration (due to the shell width and the angular
spreading times) plus the scattering duration. This total duration is
then used as a scale parameter for the pulse duration (the same as
$\alpha$ in Eq.~\ref{eq:lci}), assuming a fixed functional shape for
the pulse\footnote{The approach of Kobayashi et al. (2002) was indeed
different, in fact they assumed that all the collision between Thomson
thick shells, typically those dominated by diffusion, do not produce
any emission but reconvert the internal energy to outflow motion.}.
On the contrary, when the diffusion time scale is comparable to or
larger then the intrinsic time scale, the shape of the pulse should be
changed, smoothing out its sharp features but leaving the long time
scale features unaffected. When simulated lightcurves are observed in
Fourier space, in both cases one has as expected that the average
frequency has decreased. However, the real scattering decreases the
mean frequency by suppressing the high frequencies, while the above
approximation rescales the whole spectrum at smaller frequencies. The
result is that, while real scattering produces a prominent break in
the spectrum, this approach leaves the spectral shape unaffected.

\section{Discussion}

How do the shot-noise theory compares to the average spectra derived
in \S~3? The observed spectra show a clear break, which we show is
strongly correlated with the variability parameter. The power-law
slope changes from $-2/3$ before the break to $\sim -2$ after the
break.

Let us for the moment neglect photon scattering. The fact that the
largest slope is $\sim-2$ suggests that an exponential shot noise
model can easily reproduce the observations, with a typical pulse
decay time longer than the rise time (Norris et al. 1996). The fact
that the low frequency slope is not flat, can be easily accommodated
invoking a power-law distribution of pulse durations
(cfr. Eq.~\ref{eq:alp})
\begin{equation}
n(\tau_d)\propto\tau_d^{-1/3}
\label{eq:sixt}
\end{equation}
The correlation between the break frequency and the variability (or
the GRB luminosity) can therefore be interpreted as a correlation
between the typical shortest relevant pulse duration and the
luminosity.  This does not mean that the lightcurve cannot have
shorter pulses. In fact, if the distribution of pulse durations has a
break, shorter pulses can be present but not contribute to the PSD.

Let us now consider the effect of photon scattering. In a very simple
scenario, all the pulses may go through a scattering screen with
opacity $\tau$ and radius $R$. In this case, a clear break should be
present in the PSD at a frequency $\nu=c/(\pi\,R\,\tau)$. It is
straightforward to show that the observed break cannot be due to
scattering but must, instead, be attributed to the intrinsic
properties of the unscattered pulses. In fact, the change of slope
before and after the break is only of unity, while scattering would
require either a jump of 2 in slope (intermediate opacity) or an
exponential cutoff ($\tau>10$). Alternatively, one can consider a
scenario in which the opacity and size of the screen are different for
different pulses. In order to be consistent with the lack of a
pronounced break in the observations, this requires that the
scattering time scale $\tau R/c$ is smaller than $\tau_d$. In this
case, the PSD of the single pulse would be affected only in the
power-law tail and its effect would not be observed in the average
PSD. In order to explain the break in the average PSD, however, one
should consider a broken power-law of pulse durations. With such
conditions one can explain the observed average power-spectrum.
However, the role of scattering is not relevant in the shape of the
PSD, and so it cannot be important in the measure of the variability
parameter $V$. It may be also envisaged a case in which $\tau
R/c\gg\tau_d$ for all pulses, with a broken power-law distribution of
the values of $\tau R$ suited to mimic the average PSD shape. In this
case, however, the lightcurve would be entirely dominated by
scattering. The pulses should have the shape of the transfer functions
$k(t)$, in contradiction with what is found in their direct analysis
(Norris et al. 1996). 

A more detailed modelling of an internal shock process should however
take into account that different pulses may not be scattered with the
same value of $\tau R/c$ since the shortest pulses are likely to be
produced closer to the engine, where the relativistic wind is more
dense and opaque. To mimic such a case, we consider a lightcurve in
which only the shortest pulses underwent scattering. In this case, the
PSD should show a steepening break followed by a flattening break when
the unscattered pulse component becomes dominant (see lowest right
panel of Fig.~\ref{fig:sim}). Again, this is not observed in real data
(see Fig.~\ref{fig:ave}). 

It may be argued that these models are too simplistic, since in real
simulations each pulse is smeared with its own value of the parameter
$\tau{}R/c$. What emerges from our analysis is that no clear signature
of scattering is present in the PSD data.  Is may still be possible
that a more detailed numerical simulation for the evolution of the
flow can include scattering in such a way that its effect is relevant
but does not produce a cutoff in the PSD. This requires however a fine
tuning of the distribution of the parameter $\tau{}R/c$ and that the
shape of at least part of the pulses is dominated by the transfer
functions of Fig.~\ref{fig:ker}. Such simulations, with a proper
treatment of the scattering, are called for if the proposed link
between scattering and variability has to be believed.

Even though scattering processes are shown to be not important in
determining the variability parameter $V$, it is still possible that
they have some relevance in shaping the spectra of GRBs. In fact, if
the scattering screen is small and thick, the cutoff frequency can be
large, in a range in which the PSD is dominated by the white
noise. This kind of scattering would not be influent for the temporal
pulse profile, but the photon energy may be shifted by
\begin{equation}
\Delta\epsilon\simeq{{\tau^2\epsilon}\over{m_e\,c^2}}\left(4kT-\epsilon\right)
\end{equation}
where $T$ is the temperature of the scattering electrons. This
mechanism for modifying the typical energy of the photons in GRB
spectra has been proposed and discussed, e.g., in Ramirez-Ruiz \&
Lloyd-Ronning (2002) and \mes~et al. (2002). In particular,
assuming that the comoving electron temperature is small and asking
that the photon energy is sizably modified
($\Delta\epsilon\sim\epsilon$) one can show that the opacity must be
\begin{equation}
\tau\sim\sqrt{{{m_e\,c^2}\over{\epsilon}}} \approx \sqrt{\Gamma}
\end{equation}
where $\Gamma$ is the bulk Lorentz factor of the flow and the rightmost
term holds if the initial observed energy of the photon was closed to
$m_e\,c^2$.

Before discussing this issue in detail, we consider that all the
computation described above are relevant if the scattering screen is
comoving with the relativistic flow. In fact, in order to preserve the
burst variability, a scattering screen at rest in the frame of the
host galaxy needs to have an optical depth much smaller than unity.
For this reason, the screens that we consider here are comoving with
the flow, they can be the same shell in which the radiation is
produced, or the total contribution of previously ejected shells.

In order for the scattering to be important in shaping the spectrum of
GRBs, one have to assume a large opacity, so that the cutoff in the
power spectrum should be exponential. Consider the bursts with largest
variability (lower left panel of Fig.~\ref{fig:ave}). The spectrum is
well described by a power-law up to observed comoving frequencies
$\nu\sim10$~Hz. We can then constrain the comoving size of the
scattering screen to be
\begin{equation}
R^\prime < 10^9 \,\Gamma^{1/2}
\label{eq:con}
\end{equation}

In the framework of internal shock, the comoving width of the shell is
given by $\Delta^\prime=r/\Gamma\sim r_0\Gamma$, where $r_0$ is the
size of the shell at the moment of the ejection. The
constraint~\ref{eq:con} implies then
\begin{equation}
r_0\lsim10^8\,\Gamma_2^{-1/2}
\end{equation}
which is consistent with the internal shock picture only if the inner
engine produces many shells, of the order of 1000 shells in a burst
lasting for 10 seconds. Without being confined in the standard
internal shock model, one can envisage a scenario in which the energy
is liberated in the fireball at radii $r<r_0 \Gamma^2$. In this case,
it must be considered that the photons will be advected by the flow,
which will become optically thin to radiation (due to the shell
expansion) in a time scale smaller than the diffusion time scale of the
photons. In this case, the break frequency would be expected to be at
$\nu\simeq c\Gamma^2/(2\pi\,R)$, which is always above the measured
10~Hz.

\section	{Conclusions}

We have computed the average power spectrum density for a set of GRB
lightcurves, divided in 6 bins according to their variability
properties and we have developed a shot noise model to be compared
with them. Photon scattering is self consistently included in the
model. We find that:

\begin{itemize}
\item The variability parameter $V$ as computed by Fenimore \& 
Ramirez-Ruiz (2000) strongly correlates with the dominant frequency of
the spectra, being defined as the maximum of the $\nu\,PSD(\nu)$
function. This frequency is relatively small, suggesting that the
luminosity of GRBs is related to the variability properties of the
lightcurve in the low frequency range ($0.1\div10$~s).
\item The average power spectra are well described by broken power 
laws with low frequency slope $\nu^{-2/3}$ and high frequency slope
$\nu^{-2}$. The break frequency is a function of the variability
parameter $V$.
\item The PSDs can be easily interpreted as due to the superposition 
of random similar shots (with a distribution of decay times and
fluences) with double exponential shape (possibly stretched, see
Norris et al. 1996). In this case the position of the break frequency
(and therefore the variability parameter $V$) is related to the 
shortest relevant pulses in the duration distribution of pulses.
\item We show that photon scattering should imprint a detectable break 
or cutoff in the PSD. The lack of such a signature makes untenable all
models in which the luminosity variability correlation is ascribed to
the smoothing of the shortest time scales in low-luminosity GRBs.
\item Under certain circumstances (a compact engine or deviations 
from the standard internal shock picture) it is possible to find
models in which photon scattering plays a relevant role in shaping the
spectra of GRBs consistent with the measured PSD.
\end{itemize}

To conclude, it must be emphasized that the results we have presented
do depend at some level on the existence of a tight correlation
between the burst variability and luminosity. In particular, all the
spectral analysis is made on the {\it rest frame frequency}, which is
calculated thanks to the redshift guessed from the above mentioned
correlation. However, the results still are valid and relevant even if
such correlation should proved not to be real. In that case, the
variability would not be related to the luminosity, but still we
should conclude that scattering is not relevant in shaping the
lightcurves of GRBs. To test this, we performed the same PSD analysis
on the {\it observed frequencies}. The average spectral shapes do
remain the same, even if the values of the brake frequencies are
different and, again, there is no sign of a cutoff that may indicate
that scattering is relevant to determine the degree of variability of
the lightcurves.

\section*{Acknowledgments}

I am indebted to Martin Rees and Luigi Stella for discussions and
suggestions that stimulated and largely improved this work. I thank
Enrico Ramirez-Ruiz for providing the variability data and for
discussion on the variability luminosity correlation. I thank Andrei
Beloborodov for discussions and for carefully reading this manuscript.

\end{document}